\documentclass[12pt,prepreint,showpacs,aps]{revtex4}
\usepackage{graphicx}

\begin{document}

\title{ Column size effects of DER fluids} 

\author{Tianyu Zhao and H. R. Ma}
\affiliation{Institute of Theoretical Physics, Shanghai Jiao 
Tong University, Shanghai, 200240, China}

\begin{abstract}
The static yield stress of dielectric electrorheological(DER) fluids
of infinite column state and chain state are calculated from the 
first principle
method. The results indicate that the column surface contributions to ER
effects is very small and both states will give
correct results to the real DER fluids.
\end{abstract}
\maketitle

In the past two decades, there has been a renewed interest in the 
researches of 
Electrorheological (ER) fluids. This is partly stimulated by 
the potential and practical applications 
of this material, and partly by the desire of understanding of the physics
of many complex systems to which the ER fluids is belong.
The ER  fluids are a class of materials whose
rheological properties are controllable by the application of an
electric field.  
It is generally made up of solid particles dispersed in a liquid.
Due to the large varieties of electric properties of both the solid 
particles and liquids, and varieties of shape, size distributions
of solid particles, 
the ER mechanism is still not well understood in general.
In this paper we will focus on the simplest type of ER fluids:
uniformly sized, solid dielectric spheres dispersed in a liquid.  Both
the solid and the liquid components are assumed to follow linear
electrostatic response under an applied field. 
This type of ER fluids is denoted as the
dielectric electrorheological (DER) fluids\cite{ma96}.  
It was predicted by Halsey et al\cite{halsey} that in the DER fluids
the solid particles aggregate to form columns under electric field,
 Based on induced dipole interactions
of solid particles, 
Tao et al\cite{tao1991} argued that the structure within columns is 
body centered hexagonal(BCT).
These predictions were confirmed by 
accurate electrostatic calculations with different 
methods\cite{ma96,Davis1,Davis2,R.Tao2,Clercx}.
From calculations of Ma et al\cite{ma96,manew}, the ground state
of the DER fluid is the one that solid spheres form infinite large
columns and the structure within the column is BCT. However, experiments
usually saw chains of solid particles under weak fields and small
columns under strong fields, the size of the columns may be
determined by the dynamic process of the aggregation and is still
poorly understood. On the other hand, first principle calculations
based on the assumption of infinite large columns gave quantitative
agreement to the well controlled experiment without adjustable 
parameters\cite{ma96}.  It was argued in \cite{ma96} based on the calculation
of electrostatic free energy that the influence of column size
to static yield stress is small. We will give, in this 
paper, a comparison  of static yield stress in two extreme cases
of smallest column size(chain state) and infinite column state. The 
results of our calculation indicate that the  column size effect
is indeed small, as asserted in \cite{ma96}.

The method of calculation in this study is the first principle method 
developed by Ma et al\cite{ma96,manew}, 
based on the Bergman-Milton representation of effective dielectric
constant of a composites, and outlined in the following.

Consider a  DER model system consisting of spherical solid particles
of radius R and dielectric constant \( \epsilon _{1} \) dispersed in a fluid
characterized by \( \epsilon_2 \).
The equation for the electrostatic 
potential \( \phi (r) \) is : 
\begin{equation} \label{eq:001}
\nabla .\left[ \epsilon ({\bf r})\nabla \phi ({\bf r})\right] =0,
\end{equation}
 where the dielectric constant 
\( \epsilon ({\bf r})=\epsilon _{2}\left[ 1-\eta ({\bf r})/s\right]  \), 
\( s=\epsilon _{2}/(\epsilon _{2}-\epsilon _{1}) \) is the only 
material
parameter in the problem, and
\( \eta ({\bf r}) \) is the characteristic function of the solid 
component, defined
as having the value 1 at those spatial points occupied by the solid particles,
and zero otherwise. Following
Bergman and Milton\cite{bergman}, the solution of Eq.(1) directly yields the 
effective dielectric constant
\( \overline{\epsilon }_{zz} \) of the system,

\begin{equation} \label{eq:002}
\overline{\epsilon }_{zz}=\epsilon _{2}\left( 1-\frac{1}{sV}
\int d{\bf r}\eta \frac{\partial \phi ({\bf r})}{\partial z}\right)
 =\epsilon_2\left(1-\sum_n \frac{f_n}{s-s_n}\right)
\end{equation}
 where \(  \phi  =\left[ 1-\Gamma /s\right] ^{-1} z  \) is
the formal solution to Eq.(1), with 

\[
\Gamma =\int dr^{'}\! \eta ({\bf r}^{'})\nabla ^{'}
G({\bf r},{\bf r}^{'})\cdot \nabla ^{'}\]
 and \( G({\bf r},{\bf r}^{'})=1/(4\pi | {\bf r}-{\bf r}^{'}|)  \) being 
Green's function
for the Laplacian operator, $V$ denotes the sample volume. $s_n$
is the eigenvalue of the $\Gamma$ operator and $f_n$ related to the 
eigenfunction of  $\Gamma$. Both $s_n$ and $f_n$ are determined
only  
by the microstructure of the material and independent of the material 
parameters.  Detailed  description
of the effective dielectric constant calculation was given 
in \cite{ma96,manew}. 

Since the free energy density of the system in the 
high-field
regime is given by \( f=-\overline{\epsilon}_{zz}E^{2}/8\pi  \), the ground
state structure is determined by maximizing \( \overline{\epsilon}_{zz} \)
with respect to the coordinates of the solid particles. For ER fluids, a great
deal of experiments and calculations show that the BCT
structure is the optimal structure. To calculate the static yield stress, one
just needs to perturb the system away from the ground state by applying a shear
deformation to the system that is perpendicular to the applied field direction.
The shear distortion consists of two parts, the $c$ axis of the BCT structure
is both tilted by an 
angel \( \theta  \)
with respect to the field as well as elongated, and the $ab$ plane is
shrunken uniformly to stabilize the system by hard core repulsion of spheres.
The volume fraction of solid
spheres in the structure is also \( \theta  \) dependent 
under shear, given
by \( p_{0}(\theta )=4\pi \cos ^{2}\theta /3(8\cos ^{2}\theta -2) \). 
The free energy density of the system can be  writes as :

\begin{equation} \label{eq:003}
f=-(\overline{\epsilon }_{cc}\cos ^{2}\theta 
+\overline{\epsilon }_{aa}\sin ^{2}\theta )\frac{E_{0}^{2}}{8\pi}
\end{equation}
where \( E_{0} \) denotes the applied electric field, along $z$ direction. 
\( \overline{\epsilon }_{cc} \), the $cc$ component of the effective
dielectric tensor of the system,  is accurately related to
the $cc$ component of  effective dielectric constant within
the tilted BCT structure, $\overline{\epsilon}^\infty_{cc}$,  by the relation 
\begin{equation} \label{eq:004}
\overline{\epsilon }_{cc}=\frac{p}{p_{0}}\overline{\epsilon }^\infty_{cc}+
(1-\frac{p}{p_{0}})\epsilon _{2}
\end{equation}
 where \( p \) denotes the volume fraction of spherical solid particles of
the system. While   \( \overline{\epsilon }_{aa} \) may be 
approximated by the Maxwell-Garnett formula\cite{garnett}: 
\begin{equation} \label{eq:005}
\overline{\epsilon }_{aa}=\frac{\overline{\epsilon }^\infty_{aa}(1+p/p_{0})
+\epsilon _{2}(1-p/p_{0})}{\overline{\epsilon }^\infty_{aa}(1-p/p_{0})
+\epsilon _{2}(1+p/p_{0})}
\end{equation}
with $\overline{\epsilon}^\infty_{aa}$ the $aa$ component of the effective 
dielectric constant within the tilted BCT structure. Here $a$ and $c$ are
the crystallography coordinates of the tilted BCT structure.

The stress-strain relation is given by \( \partial f/\partial \theta  \).
To calculate the stress from effective dielectric constant, the effective
dielectric constant at a mesh of strain($\theta$) was generated, then
the differentiation was numerically calculated by
 the third order spline interpolation method.
The static yield stress is defined as the maximum stress
as function of strain. 

\begin{figure}
{\includegraphics[width=0.5\linewidth,angle=0]{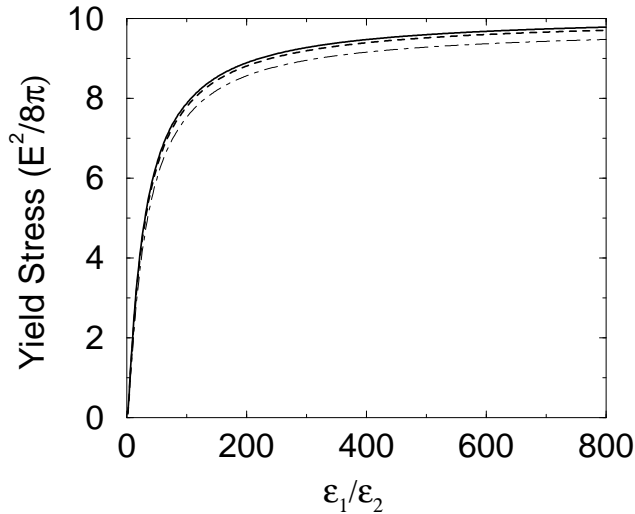} }%
\caption{Comparison between the static yield stress of 
chain states of  simple square lattice(dashed
line), the expanded BCT(solid line) and the 
       triangular lattice(dot-dashed line),  
 $\delta= 0.01$.}
\label{fig:000}
\end{figure}

In the case of chain state, the starting structure is assumed to be 
single aggregated chains
form a regular two dimensional(2D) lattices, the 2D lattice constant is 
chosen according
to the prescribed over all volume fraction $p$. By shearing the system,
the spheres within each chain are assumed uniformly separated.  
$\overline{\epsilon}_{cc}$ and $\overline{\epsilon}_{aa}$ are calculated
by method described above. It should be noted that the distorted chain
states are unstable to the formation of small aggregated
short chains by electrostatic interaction.
 However, we use these seemingly unphysical state for the
following two reasons. First, if we use small columns of 2 or more chains,
the distorted structure can be stabilized by hard core repulsions of spheres,
and the single chain is a limiting case of small columns.
Second,  we study here only the electrostatic
force after distortion, we may think that the system is dielectric
particles trapped in a gel so that the structure is stabilized. 
Since there is no phase separation in the chain
state, the average procedure described by Eq. (\ref{eq:004}) 
and Eq. (\ref{eq:005})
are not needed. We did calculations with  
simple square lattice, the
square lattice with neighbour  chains shifted by the sphere radius (the
expanded BCT) and the triangular lattice. 
The results are showing in figure \ref{fig:000}.
All lattices basically give the same results, the largest 
difference is about 3.1\%.

In our calculation, we choose \( \epsilon _{2} \) to be 2.5 
while \( \epsilon _{1} \)
ranging from 2.5 to 2000. The static dielectric constant depends not only on
\( \epsilon _{1} \), \( \epsilon _{2} \) and the volume fraction, but also 
sensitively on
the closest distance of two nearest spheres, characterized by \( \delta  \), 
related to the center to center distance of two nearest spheres $d$
by
$$
d=2R(1+\delta)$$.

\begin{figure}
{\includegraphics[width=0.5\linewidth,angle=0]{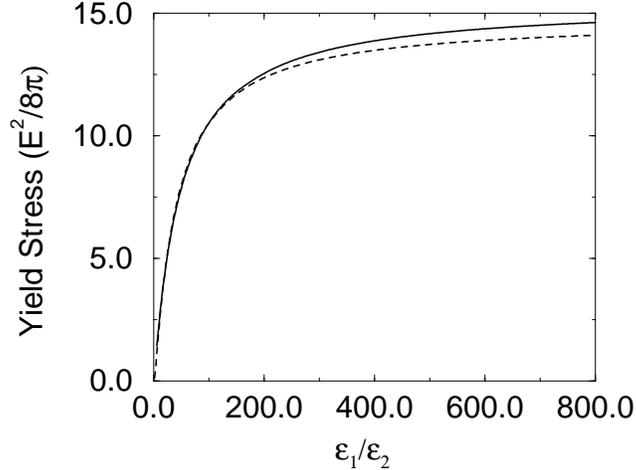} }%
\caption{Comparison between the static yield stress of 
infinite column state(solid line) and
the expanded BCT chain state(dashed line). 
The  volume fraction of the system is  0.2,
 $\delta=0.005$.}
\label{fig:001}
\end{figure}

\begin{figure}
{\includegraphics[width=0.5\linewidth,angle=0]{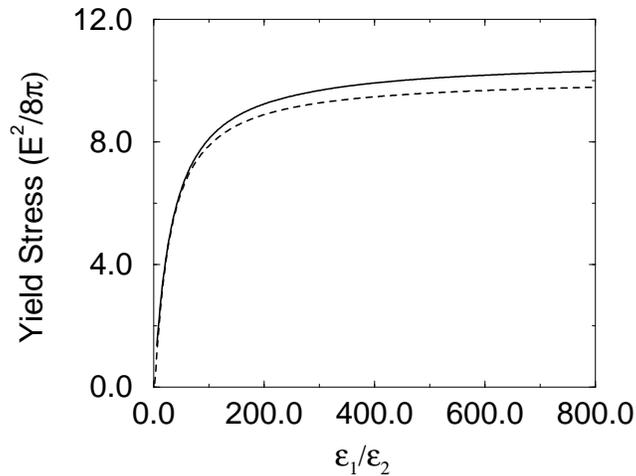} }%
\caption{Same as figure \ref{fig:001}, with $\delta=0.01$}
\label{fig:003}
\end{figure}

Figure \ref{fig:001} and figure \ref{fig:003} are plots of the 
calculated static yield stresses for the two cases with different $\delta$. 
As expected, the static yield stress of the infinite column state is always 
larger then the chain state. The important part of the figures is that
the dependence of static yield stress on the ratio of the 
dielectric constant $\epsilon_1/\epsilon_2$ is very similar. 
The difference between the two is very small. The largest
difference was found on the case of smallest $\delta$ and
largest dielectric constant mismatch, which is only 
3.75\%.

As pointed out in \cite{ma96}, the surface energy of columns in DER fluids
is very small compared to the bulk electrostatic free energy. Our 
calculation indicates that not only the surface energy is small,
its contribution to the static yield stress is also very small. 
Attempts to improve the electrorheological effects by increase
the column size is extremely limited, at least in the DER case.
This is why the theory give excellent agreement in a controllable,
pure DER system as reported in \cite{ma96}.
In a recent publication, Rong et al.\cite{rong} reported their experiment
on the large enhancement of static yield stress of magnetorheological(MR)
fluid under compression. They related this effect to the observed
column size enlargement after compression under magnetic field. 
Based on the calculation and discussions given here, we believe that
the magnetic force is not very sensitive to the structure. In any
ER and MR experiment, the 
observed static yield stress of ER fluids are contributions
both from electrostatic(or magnetic) interactions and other sources.  
When spheres are tightly packed by compression under magnetic field,
the dry friction may play
a very important rule in the static yield stress.

In this work we discussed  two extreme conditions, the infinite column
state and the  chain state. While ER fluids with thick irregular columns
arranged randomly is the observed situation in most experiments, 
which may be viewed as intermediate to the two cases studied here.
Since our results for the two extreme cases are very close to 
each other in a large range of other parameters, it is a good
approximation to use both cases in explaining  experiment
results  of DER fluids. 
This work is supported by 
The Research Fund for the Doctoral Program of Higher Education
and  National Natural Science Foundation of China.
We thank the referee to point out reference\cite{rong}.

\end{document}